\documentclass[preprint,12pt]{elsarticle}

\usepackage[graphicx]{realboxes}
\usepackage{varwidth}

\usepackage{amssymb}

\usepackage{lineno}

\biboptions{sort&compress}

\usepackage{xcolor}

\newcommand\ApJ{Astrophys.J. }
\newcommand\AaA{Astron.Astrophys. }
\newcommand\EPJ{Eur.Phys.J. }
\newcommand\nima{Nucl.Instrum.Methods A }
\newcommand\nim{Nucl.Instrum.Methods }

\newcommand\beq{\begin{equation}}
\newcommand\beql[1]{\begin{equation} \label{#1}}
\newcommand\eeq{\end{equation}}
\newcommand\ben{\begin{eqnarray}}
\newcommand\een{\end{eqnarray}}
\newcommand\bea{\begin{array}}
\newcommand\eea{\end{array}}
\newcommand\bem{\begin{displaymath}}
\newcommand\eem{\end{displaymath}}

\newcommand\eqa[1]{Eq.~(\ref{#1})}
\newcommand\eqb[1]{Eqs.~(\ref{#1})}
\newcommand\eqc[1]{(\ref{#1})}

\newcommand\fig[1]{Fig.~\ref{#1}}

\newcommand\tab[1]{Table~\ref{#1}}

\newcommand\sct[1]{Section~\ref{#1}}
\newcommand\sctw[2]{Sections~\ref{#1} and~\ref{#2}}

\newcommand\sctf[4]{Sections~\ref{#1},~\ref{#2},~\ref{#3} and~\ref{#4}}
\newcommand\app[1]{\ref{#1}}

\newcommand\qqb{\qquad}
\newcommand\qqc{\qquad \qquad}

\newcommand\wse{\vspace{-3.5cm}}

\newcommand\Ssum[2]{\sum \limits_{#1}^{#2}}

\newcommand\Iint[2]{\int \limits_{#1}^{#2}}
\newcommand\dif[1]{{\rm \,d}{#1}}
\newcommand\sgn{{\rm sgn}}

\newcommand\non{n_{\rm on} }
\newcommand\noff{n_{\rm off} }

\newcommand\mon{\mu_{\rm on} }
\newcommand\moff{\mu_{\rm off} }

\newcommand\ms{\mu_{\rm s} }
\newcommand\mb{\mu_{\rm b} }

\newcommand\Bii{{\rm Bi} }
\newcommand\NBii{{\rm NBi} }

\newcommand\Poo{{\rm Po} }
\newcommand\Gaa{{\rm Ga} }

\newcommand\fGa{f_{\rm Ga} }

\newcommand\PPo{P_{\rm Po} }

\newcommand\SLM{S_{\rm LM} }
\newcommand\SBa{S_{\rm B}}

\newcommand\IIp{I}

\newcommand\delr{\delta_{+} }
\newcommand\dell{\delta_{-} }

\newcommand\delrr{\delta^{+}_{+} }

\newcommand\fdel{f_{\delta} }
\newcommand\fdelp{f^{+}_{\delta} }

\newcommand\fdelb{h_{\delta} }
\newcommand\fdelbp{h^{+}_{\delta} }

\newcommand\fdelr{g_{\delta'} }
\newcommand\fdels{f_{\delta}' }

\newcommand\Pm{P^{-} }
\newcommand\Pp{P^{+} }

\newcommand\Ppb{R^{+} }

\newcommand\aon{\frac{\alpha}{1+\alpha} }
\newcommand\aoff{\frac{1}{1+\alpha} }
\newcommand\alal{\frac{1+\alpha}{\alpha} }
\newcommand\alin{\frac{1}{\alpha} }

\newcommand\ssp{s_{p} }
\newcommand\ssq{s_{q} }
\newcommand\gp{\gamma_{p} }
\newcommand\gq{\gamma_{q} }
\newcommand\gqa{\frac{\gamma_{q}}{\alpha} }
\newcommand\gqb{\left( \frac{\gamma_{q}}{\alpha} \right) }

\newcommand\gpqb{\gamma_{p} + \frac{\gamma_{q}}{\alpha} }
\newcommand\gpqc{\frac{\alpha \gamma_{p}}{\gamma_{q}} }

\newcommand\rrha{\frac{1}{1+\rho} }
\newcommand\rrhb{\frac{\rho}{1+\rho} }

\journal{Nuclear Instruments and Methods A}

\begin{document}

\begin{frontmatter}

\title{On Bayesian analysis of on--off measurements}


\author[nos01]{Dalibor Nosek}
\ead{nosek@ipnp.troja.mff.cuni.cz}
\author[nos02]{Jana Noskov\'a}
\address[nos01]{Charles University, Faculty of Mathematics and Physics,
Prague, Czech Republic}
\address[nos02]{Czech Technical University, Faculty of Civil Engineering,
Prague, Czech Republic}

\begin{abstract}
We propose an analytical solution to the on--off problem 
within the framework of Bayesian statistics. 
Both the statistical significance for the discovery of 
new phenomena and credible intervals on model parameters 
are presented in a consistent way.
We use a large enough family of prior distributions 
of relevant parameters. 
The proposed analysis is designed to provide Bayesian 
solutions that can be used for any number of observed on--off 
events, including zero.
The procedure is checked using Monte Carlo simulations.
The usefulness of the method is demonstrated on examples
from $\gamma$--ray astronomy. 
\end{abstract}


\end{frontmatter}


\section{Introduction}
\label{S01}

We consider an on--off experiment that is designed for counting 
two classes of events, source and background events, the type of 
which cannot be distinguished in principle.
These events are registered in two disjoint regions characterized 
by some sets of coordinates.
We deal with small numbers of events the detection rates of which are 
modeled as independent Poisson processes with unknown means. 

The problem of the on--off measurement we want to solve is whether 
the same emitter with a constant but unknown intensity is responsible 
for the observed counts in both studied regions.
Any inconsistency between the numbers of events collected in these 
regions, when they are properly normalized, then speaks in favor 
of the predominance of a source producing more events in one 
of the explored region over the other. 

Techniques for addressing these issues from a classical point of view 
have been presented in the literature.
The likelihood ratio test together with the Wilks' theorem~\cite{Wil01} 
are often utilized to characterize asymptotically the level of agreement 
between the data and the assumption of new 
phenomena, see e.g. Refs.~\cite{Lim01,Cou01,Cow01}. 
A widely discussed problem is of how to establish upper bounds of 
the source intensity for small numbers of detected 
counts~\cite{Cow01,Fel01,Cou02,Rol01}. 
A Bayesian solution of the on--off problem was proposed in 
Refs.~\cite{Hel01,Hel02} when analyzing multichannel spectra in
nuclear physics.
More general solutions can be found in Refs.~\cite{Pro01,Pro02,Gil01,Gre01}.
Very recently, two specific Bayesian solutions to the on--off problem 
have been presented~\cite{Kno01,Cas01}. 

In this study, we focus on how to confirm the presence of a weak 
source and on the determination of credible intervals for its 
intensity at a given level of significance. 
We address these issues from a Bayesian point of view.
Our original intent is to provide different insights pertaining 
to the on--off problem that benefit from its simplicity. 
We do not follow Bayesian alternatives to classical hypothesis testing 
that deal with priors for models (hypotheses) and compare competing 
models in terms of Bayes factors. 
In our concept, the plausibility of possible models for the on--off 
data is assessed by parametrizing the space of models using 
a suitable parameter. 
Here we use the difference between the on--source and background 
means inferred from the on--off measurement, treating these 
quantities as random variables within the Bayesian setting.
Information on the various aspects of the investigated phenomena 
are accessible in the posterior distribution of this difference.

Our {\it \`a priori} knowledge about the underlying Poisson
processes is consistently improved by using only the on--off data 
without any external assumptions. 
By finding the extent to which the on--source Poisson mean 
is greater or less than the background one, this option allows us 
to obtain a new well--reasoned formula for the Bayesian 
probability of the source presence in the on--source zone.
As other Bayesian approaches to the on--off problem, we also 
receive solutions in the case of small numbers, including 
the null experiment or the experiment with no background, 
when classical methods based on the asymptotic properties of 
the likelihood ratio statistic~\cite{Lim01,Cou01,Cow01} are 
not easily applicable.
In addition, our strategy allows us to establish limits on 
parameters for the processes that are responsible for 
the observed phenomena. 
There are no problems with the discreteness of counting 
experiments or with unphysical likelihood estimates,  
see e.g. Refs.~\cite{Fel01,Cou02,Rol01}.
We provide credible intervals that are likely to be very similar 
to the way in which the experimental results are usually communicated. 

The proposed method is particularly suitable when dealing with 
peculiar sources whose observational conditions cannot be set in 
an optimal way. 
Examples include transient $\gamma$--ray sources and gamma--ray 
bursts when searching for an accompanying signal on targets 
in their space--time vicinity. 
This approach can also advantageously be used to assess possible 
sources of charged cosmic rays with characteristics hypothesized 
in previous measurements.

The structure of the paper is as follows. 
The essential features of our approach are described in detail 
in~\sct{S02}.
We present and discuss general formulae for assessing the existence  
of the source and for estimating its activity.
Particular attention is paid to cases with uninformative priors.
Examples are presented and discussed in~\sct{S03}. 
The paper is concluded in~\sct{S04}.

\section{Bayesian solutions to on--off problem}
\label{S02}

In a typical on--off analysis, a measurement of a physical quantity 
of interest is set by comparing the number of events $\non$ recorded 
in a signal (on--source) region, where a source is expected, with 
the number of events $\noff$ detected in a reference (off--source) 
region.
The on--off data, especially when only a few events are recorded, 
are modeled as discrete random variables generated in two independent 
Poisson processes with unknown on-- and off--source means, 
$\mon$ and $\moff$, 
i.e. $\non \sim \Poo(\mon)$ and $\noff \sim \Poo(\moff)$\footnote{
Throughout this study, we use the same symbols for random variables 
and their sample values.}, respectively.

The relationship between both the on-- and off--source regions 
is given by the ratio of on-- and off--source exposures, 
by the on--off parameter $\alpha > 0$. 
This parameter includes, for example, the ratio of the observational 
time for the two kinds of events and the ratio of their collecting 
areas modified by corresponding experimental efficiencies.  
Its value is assumed to be known from the experimental details. 
It can be estimated from additional measurements or extracted from 
a model of the detection.
Relying upon that, the unknown mean of background counts in 
the on--source region is $\mb = \alpha \moff$.

In our treatment, the on--off problem consists in the assessment 
of the relationship between the two unknown on-- and off--source 
means, $\mon$ and $\moff$.
To solve this task we utilize Bayesian reasoning. 
It is worth pointing out that we do not use often adopted 
scheme, whereby the source and background parameters, that are 
responsible for observed on--source counts, are chosen as 
the basic independent variables, see e.g.~\cite{Gre01,Kno01,Cas01}.
We proceed quite differently. 
In our concept, the Bayesian inference is applied to improve 
our knowledge about the observed phenomena without any external 
assumptions about the relationship of the underlying processes.
We do not compare models with and without a source in 
the on--source zone, as usually proposed, i.e. no hypotheses 
about the source presence are tested nor Bayes factors for 
on--source model selection are examined. 
In addition to our {\it \`a priori} notion derived from our previous 
experience or just selected with respect to our ignorance, for example, 
we use only experimental data in order to assess whether a source 
may be identified in the on--source region.
We also show how new information may be incorporated in our treatment. 
This scheme is not only backed by a compelling statistical 
motivation, but also fairly simple to implement, yet sufficiently 
general. 
Nonetheless, our results may deviate from the results obtained 
under the assumptions used in other Bayesian inference 
methods aiming to analyze the on--off 
problem~\cite{Pro01,Pro02,Gre01,Gil01,Kno01,Cas01}.
Based on that, our findings are to be interpreted differently
in some cases.

In the first step of our analysis, we focus on what kind 
of information about the on-- and off--source means can be obtained 
from the on--off measurements provided that observed counts in 
both zones follow the Poisson distribution. 
Since we have no {\it \`a priori} knowledge whether or in what way 
these means are related, we examine their independent prior 
distributions.
Using the on--off data and our prior information about the on-- 
and off--source means, we derive their marginal posterior distributions. 
These distributions summarize our state of knowledge and remaining 
uncertainty about the on--source mean $\mon$ and separately about 
the off--source mean $\moff$, given the data.
Thus, the probability that the on--source mean acquires a certain 
value is given without reference to values of the off--source mean,
and vice versa.

In the second step, we compare information we have about 
both inferred means.
Using a known on--off parameter $\alpha$, we normalize the off--source 
mean in order to obtain a parameter that corresponds to 
the on--source exposure, i.e. 
we construct the marginal distribution for the parameter 
$\mb = \alpha \moff$. 
Then, we determine which of the observed on--off processes 
is more substantial without any assumptions about the relationship 
of the underlying processes. 
In order to get the most unbiased value of the source probability, 
we assume a maximally uninformative joint distribution of the on--source 
and background means, given the on--off data. 
For this purpose, we examine the product of their marginal posterior 
distributions, as dictated by the principle of maximum entropy, and 
construct the distribution of their difference $\delta = \mon - \mb$
with a real valued domain.
The probability with which $\delta > 0$, as inferred from 
this distribution, tells us what is the probability that a larger 
intensity is detected in the on--source region than expected from 
the off--source measurement.

In more detail, the posterior distribution of the difference $\delta$
allows us to decide whether a source is or is not present in the on--source 
zone. 
The presence of a source in the on--source region is 
validated if the on--off data prefers $\delta > 0$ at a given level 
of significance.
On the other hand, if the data indicates that $\delta \le 0$ at a chosen 
level of significance, we state that a source is not present in 
the on--source region. 
Instead, we infer that the data suggests more activity 
observed in the off--source region than in the on--source zone.

Although, in the Bayesian context, the assumption of a non--negative 
source rate ($\delta \ge 0$) is often taken into account by noting 
that its negative values are 
unphysical~\cite{Pro01,Pro02,Gil01,Gre01,Kno01,Cas01}, 
we initially do not require that a source is present in 
the on--source region.
In this step, our choice of the on-- and off--source region 
is regarded as purely formal considering the fact that it 
is not clear in advance what the data will reveal. 
This feature brings our analysis closer to widely used 
classical methods that strive for knowledge about the source 
activity without constraints on the properties of the underlying 
processes and benefit just from the maximum likelihood estimates 
of relevant parameters, see e.g. 
Refs.~\cite{Lim01,Cou01,Cow01,Fel01,Cou02,Rol01}.
In this sense, we adopt more precise information about 
the same parameters including their uncertainties. 
This information is contained in their marginal posterior 
distributions obtained with the help of the on--off data using 
the Bayesian reasoning.
We end up with simple expressions that, in agreement with our 
initial knowledge and without any additional assumptions, 
describe what can be learned about the source presence in 
the on--off experiment.

In the final step of our analysis, we focus on the important case 
when it is known before the measurement is carried out that a source 
may be present only in the on--source region.
Under this condition, our initial ignorance about the relationship 
between the on-- and off--source processes is updated.
With the prior requirement that only the joint distribution 
of the on--source
and background means satisfying $\mon \ge \mb$ ($\delta \ge 0$)
is admissible, the dependence between on-- and off--source 
processes reappears in the posterior probability distribution of 
the non--negative difference.
Using this distribution, we finally obtain credible intervals 
or upper bounds of the intensity of a possible source that is 
expected in the on--source zone.
By their construction, these limits are to be non--negative. 
In special cases, we obtain the posterior distributions of the 
non--negative difference that are in agreement with 
the distributions of the source rate that have been derived using 
a joint prior distribution with dependent on-- and off--source 
parameters ( $\mon \ge \alpha \moff$, $\moff > 0$)
within different Bayesian 
approaches~\cite{Hel01,Hel02,Pro01,Pro02,Gre01,Kno01}.

\subsection{On--off means}
\label{S02a}

We consider that $\non$ and $\noff$ counts were registered 
independently in the on-- and off--source regions, respectively. 
We treat the on-- and off--source data separately on 
an equal footing 
and construct the marginal posterior distributions of
the on-- and off--source means $\mon$ and $\moff$. 
This way, information about the on--source mean contained 
in its posterior distribution is given without reference 
to what is known about the off--source mean, and vice versa.

For both these random variables, we adopt a sufficiently large 
family of conjugate prior distributions for the Poisson 
likelihood function.
Specifically, we assume that the prior probability distribution 
of the on--source mean is $p(\mon) = \fGa(\mon \mid \ssp,\gp-1)$ and, 
in a similar way, the prior distribution of the off--source 
mean is $p(\moff) = \fGa(\moff \mid \ssq,\gq-1)$, where 
the Gamma distribution
$\fGa = \fGa(\mu \mid s,\lambda)$ is introduced in~\app{App01}.
The prior shape parameters, $\ssp >0$ and $\ssq > 0$, and the prior 
rate parameters, $\gp > 1$ and $\gq > 1$, characterize our information 
about the on-- and off--source zones before the measurement began.

Here we allow that the prior parameters for the on-- and off--source 
means can acquire different values. 
This freedom is due to the fact that we can have in principle 
different initial knowledge about the on-- and off--source zone. 
Such informative prior distributions express our specific knowledge 
of the examined parameters that may be taken from other 
experiments or from theoretical considerations, for example.
On the other hand, when no such input information is available,
the use of uninformative prior distributions (small values 
of the prior parameters, e.g. $0 < s \le 1$ and $\gamma \to 1$) 
typically yields results which are not too different from 
the results of conventional statistical analysis.

The posterior distributions of the means $\mon$ and $\mb = \alpha \moff$, 
given the on--off data, $\non$ and $\noff$, then take again the form 
of the Gamma distribution, see~\app{App01}.
In particular, we have $\mon \sim \Gaa(p,\gp)$ for the posterior 
distribution of $\mon$ and $\mb \sim \Gaa(q,\gqa)$ for the posterior 
distribution of $\mb$. 
Here, the shape parameters, $p = \non+\ssp$ and $q=\noff+\ssq$, 
include, except of the prior input ($\ssp$ or $\ssq$), also 
the information acquired from the on--off measurement 
($\non$ or $\noff$).
The rate parameters of the posterior distributions of the on-- 
and off--source means are given by the prior rates $\gp$ and $\gq$, 
respectively.
The rate parameter of the posterior distribution of the background 
mean $\mb$ is modified according to the exposures of the on-- and 
off--source zones as expressed by the on--off parameter $\alpha$. 

\subsection{Difference of on--off means}
\label{S02b}

In the context of a single on--off measurement, we address 
a question of what we are able to learn about the relationship
of the underlying Poisson processes that generate the observed 
on-- and off--source counts.
Since there is no other relevant information, we assume that 
the joint probability distributions of the involved 
on--source and background means is given by the product 
of their marginal posterior distributions, as it results from 
maximizing missing information.

With the marginal posterior distributions of the on--source 
and background means, $\mon$ and $\mb = \alpha \moff$, derived from 
the on--off data (see~\sct{S02a}), we arrive at the first important 
result of our study.
Under the transformation $\delta = \mon - \mb$ while keeping $\mb$ 
unchanged, with the Jacobian $J = 1$, and then marginalizing over $\mb$, 
we obtain after the standard calculations the probability distribution 
of the difference of these two unknown means 
(to simplify the notation we denote 
$\fdel(x) = \fdel( \delta=x \mid \non, \noff, \IIp)$ where 
$\IIp = (\ssp, \ssq, \gp, \gq, \alpha)$ stands for prior 
information)\footnote{
In the following, the explanatory variable $x$ in the probability 
distribution denotes the values which the corresponding random variable 
may acquire, in the sense that, for example, the probability 
$P( x < \delta \le x + \dif{x} ) = \fdel(x) \dif{x}$.}
\beql{B01}
\fdel(x) = 
\frac{\gp^{p} \gqb^{q}}{\Gamma(p)} \
e^{-\gp x} x^{p+q-1} \
U(q,p+q, \eta x), 
\qqc
x \ge 0,
\eeq
\beql{B02}
\fdel(x) = 
\frac{\gp^{p} \gqb^{q}}{\Gamma(q)} \
e^{\gqa x} (-x)^{p+q-1} \
U(p,p+q,-\eta x), 
\qqc
x < 0.
\eeq
Here $p = \non+\ssp$ and $q=\noff+\ssq$ where 
$\ssp >0$, $\ssq > 0$, $\gp > 1$ and $\gq > 1$ are the prior
parameters, $\eta = \gpqb$, 
$\Gamma(a) = \Iint{0}{\infty} e^{-t} t^{a-1} \dif{t}$ 
stands for the Gamma function and 
\beql{B03}
U(a,b,z) =  \frac{1}{\Gamma(a)}
\Iint{0}{\infty} 
e^{-z t} t^{a-1} (1+t)^{b-a-1} \dif{t},
\eeq
is the integral representation of the Tricomi confluent hypergeometric 
function~\cite{Olv01}. 
The probability distribution written in~\eqb{B01} and~\eqc{B02} is our 
full inference about the difference of the two unknown means $\mon$ 
and $\mb = \alpha \moff$ given the on--off data.
This solution is maximally noncommittal with respect to unavailable 
information about the relationship between these means. 
Note that, by definition, the domain of the new random variable 
$\delta = \mon - \mb$ is not limited and this difference may take 
all real values.

In practical applications, the integrals in~\eqa{B03} can be 
calculated numerically.
The saddle point approximation can be used with a good precision
if the parameters $p > 1$ and $q > 1$.
Analytic expressions can be obtained when selecting particular 
parameters of the prior distributions.

In some cases, it may be preferred to work with integer values of 
the parameters $p$ and $q$.
Then, the Tricomi confluent hypergeometric function in~\eqa{B03} may 
be after some calculations expressed as a finite series ($a,b \in {\rm N}$)
\beql{B04}
U(a,b,z) = z^{1-b} \ (b-a-1)! \ Q(a,b,z), 
\eeq
where
\beql{B05}
Q(a,b,z) = 
\Ssum{i=0}{b-a-1} {b-i-2 \choose b-a-i-1} \frac{z^{i}}{i!}.
\eeq
Straightforward calculations then give ($p,q \in {\rm N}$)
\beql{B06}
\fdel(x) = 
\frac{\gp^{p} \gqb^{q}}{\eta^{p+q-1}} \
e^{-\gp x} \
Q(q,p+q,\eta x), 
\qqc 
x \ge 0,
\eeq
\beql{B07}
\fdel(x) = 
\frac{\gp^{p} \gqb^{q}}{\eta^{p+q-1}} \
e^{\gqa x} \
Q(p,p+q,-\eta x), 
\qqc 
x < 0.
\eeq
Special examples are $\ssp = \ssq = 1$ or a limiting 
case when $\ssp = \ssq \to 0$ (for $\non > 0$ and $\noff > 0$). 

Notice that, if $\gp = \gq$, any probabilistic conclusion based 
on the distribution of the difference is independent of the choice 
of the common prior rate. 
This property is confirmed when integrating the distribution for 
the difference over an arbitrary interval.
Indeed, after a suitable transformation of variables it turns out 
that the result of integration, and the cumulative distribution 
function in particular, depends only on the ratio $\frac{\gp}{\gq}$. 

It is also worth mentioning the following property of 
the difference of the on--source and background parameters.
Let us define a new random variable $\delta' = -\delta / \alpha$.
Then, it is easy to show that its distribution function, $\fdelr(x)$, 
satisfies $\fdelr(x) = \fdels(x)$ where $\fdels(x)$ denotes the distribution 
function of the difference $\delta = \mon - \alpha \moff$, 
as given in~\eqb{B01} and~\eqc{B02}, 
that is obtained under the transformation
$(p,q,\gp,\gq,\alpha) \to (q,p,\gq,\gp,\alin)$.
Stated differently, when the on-- and off--source regions 
are exchanged,
i.e. $(\non, \noff, \alpha) \to (\noff,\non,\alin)$,
and, accordingly, prior information is exchanged, 
$(\ssp,\ssq,\gp,\gq) \to (\ssq,\ssp,\gq,\gp)$,
then the resulting distribution function describes the difference 
$\delta' = \moff - \mon / \alpha$.
Thus, any imbalance between the involved regions leads to 
the same statistical conclusion irrespective what is 
the reference region.
Any excess of counts in one of these zones that suggests the source 
presence therein is equivalently described as an unknown process 
that reduces the number of events in the complementary region.

From this point of view, it is worth bearing in mind that 
other classical test statistics possess the same property.
For example, the asymptotic Li--Ma significance, see Eq.(17) 
in Ref.~\cite{Lim01}, is in this sense invariant under 
the transformation $(\non, \noff, \alpha) \to (\noff,\non,\alin)$.
In the binomial treatment~\cite{Cou01}, the binomial $p$--value 
for a deficit of counts in the new on--source region is equal 
to the $p$--value for an excess of counts in the original 
on--source zone.
Also the asymptotic binomial formula for the source detection, 
see e.g. Eq.(9) in Ref.~\cite{Lim01}, has the same characteristics.
In a similar manner, when the on-- and off--source regions are 
exchanged, it is easy to show that the transformed profile 
likelihood ratio, see e.g. Ref.~\cite{Rol01}, provides asymptotic 
confidence intervals for 
$\delta' = -\delta / \alpha = \moff - \mon / \alpha$.

\subsection{Source detection}
\label{S02c}

With the posterior probability distribution of the difference 
we compare the involved on--source and background means.
The Bayesian probability that the source is not present 
in the on--source region corresponds to the non--positive 
difference of the on--source and background means.
It is obtained by integrating the probability distribution of 
the difference $\delta$ given 
in~\eqa{B02} for $\mon \le \mb$, i.e. $\delta \le 0$.
After straightforward calculations we get the second important 
result of this study.
The Bayesian probability of the absence of a source in 
the on--source region takes a simple form
\beql{B08}
\Pm = P(\delta \le 0) = \Iint{-\infty}{0} \fdel(x) \dif{x} = 
I_{\rrhb}(p,q),
\eeq
where $\rho = \gpqc$, $I_{x}(a,b)$ denotes the regularized incomplete 
Beta function that is determined by  
$B(a,b) I_{x}(a,b) = B_{x}(a,b)$ where 
$B_{x}(a,b) = \Iint{0}{x} t^{a-1} ( 1 - t)^{b-1} \dif{t}$ is the
incomplete Beta function and 
$B(a,b) = B_{1}(a,b) = \frac{\Gamma(a) \Gamma(b)}{\Gamma(a+b)}$ 
denotes the Beta function~\cite{Olv01}.
Obviously, the source is observed in the on--source region with 
the Bayesian probability
\beql{B09}
\Pp = P(\delta > 0) = 1 - \Pm = I_{\rrha}(q,p).
\eeq

For practical reasons, we also define the Bayesian significance 
by $\SBa = \Phi^{-1}(\Pp)$ where $\Phi = \Phi(x)$ is the cumulative
standard normal distribution.
This significance corresponds to the number of standard deviations 
from a hypothesized value in a classical one--tailed test with 
a normal distributed variable~\cite{Cou01}. 
We use notation in which a negative value of this significance 
indicates that the absence of a source in the on--source region 
is more likely than its presence therein, i.e. if $\Pm > 0.5$.

The result written in~\eqa{B09} represents the Bayesian
probability of the source hypothesis, given the on--off data and 
our prior knowledge of the underlying processes.
It allows us to assess the extent to which the processed data 
is indicative of the source of events. 
This approach differs from the classical concept designed 
to measure the exceptional nature of the on--off data with respect 
to the background model.
Our determination of the probability of the source model also 
differs from the Bayesian strategy based on the initial 
premise of the non--negative source rate ($\mon \ge \mb$), see our 
discussion on special cases in~\sctw{S02db}{S02dc}. 
In a sense, by using a wider range of alternative models 
($\mon > 0$ and $\moff = \alpha \mb > 0$), 
our approach can yield more robust information.

The interpretation of the probability of the source presence 
in the on--source zone is valid only if the on--off experiment 
is well designed in the sense that only background counts are recorded 
in the off--source zone, the corresponding background applies in 
the on--source region where an extra source may be present.
Nonetheless, if it is not the case and, for example, an unknown source
is present in the off--source zone or the on--source region is shielded 
due to an unknown process, the resultant probabilities apply as well, 
but they should be assigned different meanings.
Naturally, the above mentioned options can not be distinguished 
in a statistical evaluation.

The Bayesian probabilities of the source absence or presence 
in the on--source region do not depend on the prior rate parameters
if $\gp = \gq$ implying $\rho = \alpha$. 
In such a case, if the parameters $p > 0$ and $q > 0$ acquire 
integer values, the result in~\eqa{B08} can be rephrased 
using the representation of the binomial distribution.
Since the probability $P(N \le q-1) = I_{\aon}(p,q)$~\cite{Olv01}
where $N$ is a binomial random variable with parameters $p+q-1$ 
and $\aoff$, i.e. $N \sim \Bii(p+q-1,\aoff)$, we have ($p,q \in {\rm N}$)
\beql{B10} 
\Pm = 
\Ssum{i=0}{q-1} {p+q-1 \choose i} 
\left( \aoff \right)^{i} \left( \aon \right)^{p+q-i-1}. 
\eeq
It gives the probability that less than $q$ events out of $p+q-1$ events 
are registered in the off--source region or, alternatively, $p$ or more 
events out of $p+q-1$ events are detected in the on--source region, 
if the null background hypothesis is true, i.e. $\mon = \mb$.

Alternatively, when the parameters $p > 0$ and $q > 0$ are integers 
and $\gp = \gq$, it also holds that the probability 
$P(N \le q-1) = I_{\aon}(p,q)$ where
$N$ is a negative binomial random variable with parameters $p$ 
and $\aoff$, i.e. $N \sim \NBii(p,\aoff)$.
Then, one easily recovers that the probability of the absence of a source 
in the on--source region is ($p,q \in {\rm N}$) 
\beql{B11} 
\Pm = 
\Ssum{i=0}{q-1} {p+i-1 \choose i} 
\left( \aoff \right)^{i} \left( \aon \right)^{p}. 
\eeq
This probability describes that less than $q$ events are registered 
in the off--source region before the chosen number of $p$ events
is detected in the on--source region, if the null hypothesis stating 
that no source is present in the on--source region is true.

The above mentioned results written in~\eqb{B10} and~\eqc{B11} hold, 
for example, for the uniform prior distributions of the on-- and 
off--source means when the prior shape parameters 
$\ssp =\ssq = 1$ or for the scale invariant prior distributions 
when $\ssp = \ssq \to 0$ (for $\non > 0, \noff > 0$), 
while $\gp = \gq \to 1$.
In both these cases, the Bayesian probability of no source in 
the on--source region is similar to the classical 
probability to reject the background hypothesis, if it is true, 
in favor of an excess of the on--source events (excess $p$--value).
Note that this $p$--value follows from the classical test of 
the ratio of two unknown Poisson means~\cite{Cou01}.

Interestingly, assuming $\gp = \gq \to 1$, the probability 
of the source absence in the on--source zone derived with 
the uniform priors ($\ssp =\ssq = 1$) is higher than the corresponding 
probability derived with the scale invariant priors 
($\ssp = \ssq \to 0$ for $\non > 0, \noff > 0$), 
i.e. $\Pm(\ssp =\ssq = 1) > \Pm(\ssp =\ssq \to 0)$,  only if 
$\alpha \non > \noff$, and vice versa.
This result is easily obtained by combining the recurrence 
relations for the incomplete Beta function~\cite{Olv01}.

\subsection{Known source}
\label{S02d}

An important case occurs if it is guaranteed with certainty 
that a source may be observed only in the on--source region. 
Then, the mean event rate in the on--source zone can only 
increase beyond what is expected from background.
Such a situation is encountered when the ability of the source 
to produce detectable events has been confirmed in previous analyses 
or deduced from theoretical considerations, for example. 
In our concept, the additional knowledge about the source,
thought of as a new piece of prior information, 
is easily incorporated into the Bayesian inference by conditioning 
on the source rate.
This modification allows us to describe the properties of 
the predefined source, thus also providing us with information 
related to its detection.

Assuming that the on--source mean is not less than the 
background one, we are now dealing the case when the processes 
generating observed counts in both zones are not independent.
For this purpose, we consider the joint prior of both means 
that is written in a separable form and supplemented with 
the condition  $\mon \ge \mb = \alpha \moff$, i.e. $\delta \ge 0$.
Under this condition, the posterior probability distribution 
of the non--negative difference is easily determined by using 
the results written in~\eqb{B01} and~\eqc{B02}. 
The resultant distribution allows us to deduce a credible 
interval or an upper bound of the source intensity, while there 
are no problems with negative limits. 
It is worth emphasizing that this fairly simple construction 
of limits is equivalent to the analysis scheme in which the model with 
non--negative source intensity ($\mon \ge \alpha \moff$) 
is examined.
Therefore, using special kinds of the prior distributions, we arrive 
to the posterior distributions of the non--negative difference 
which agree with the corresponding posterior distributions of the
source intensity obtained in other Bayesian approaches, 
see~\sctf{S02da}{S02db}{S02dc}{S02dd}.

When one is concerned with the non--negative source rate, 
the corresponding probability distribution is derived 
under the condition of non--negative values of the difference 
of the on--source and background means, 
i.e. $\mon \ge \mb = \alpha \moff$ implying $\delta \ge 0$.
The distribution of the non--negative difference then follows 
from~\eqa{B01}.
Another important result of our analysis that includes several 
previously derived results~\cite{Hel01,Hel02,Pro01,Pro02,Gre01,Kno01} 
is
\beql{B12}
\fdelp(x) = 
\frac{\gp^{p} \gqb^{q} \Gamma(q)}{\Gamma(p+q) B_{\rrha}(q,p)} \ 
e^{-\gp x} x^{p+q-1} \
U(q,p+q, \eta x), \qqb
x \ge 0,
\eeq 
where we introduced the conditional distribution
$\fdelp(x) = \frac{ \fdel(x)}{\Pp}$ for $x \ge 0$, 
$\Pp = 1 - \Pm$ is the Bayesian probability that the source 
is present in the on--source region, as given in~\eqa{B09},
$\eta = \gpqb$ and $\rho = \gpqc$.

In particular, if the parameters $p > 0$ and $q > 0$ acquire integer 
values, the probability distribution of the non--negative difference 
of the on--source and background means is obtained from~\eqa{B06} 
($p,q \in {\rm N}$)
\beql{B13}
\fdelp(x) = 
\frac{\fdel(x)}{\Pp} = 
e^{-\gp x} \frac{Q(q,p+q, \eta x)}{Q_{\gp}^{+}(q,p+q,\eta)}, 
\qqb 
x \ge 0,
\eeq
where the function $Q(a,b,z)$ is given in~\eqa{B05} and $\Pp$ is 
the probability that the source is present in the on--source region 
written 
\beql{B14}
\Pp = 
\frac{\gp^{p} \gqb^{q}}{\eta^{p+q-1}} \ 
Q_{\gp}^{+}(q,p+q,\eta),
\eeq
where ($a,b \in {\rm N}$)
\beql{B15}
Q_{\gp}^{+}(a,b,z) = 
\Iint{0}{\infty} Q(a,b,z x) e^{-\gp x} \dif{x} = 
\Ssum{i=0}{b-a-1} {b-i-2 \choose b-a-i-1} \
\frac{z^{i}}{\gp^{i+1}}.
\eeq

\subsubsection{Scale invariant priors}
\label{S02da}

The scale invariant prior for a non--negative random variable corresponds 
to a uniform prior of its logarithm.
In our treatment, such prior distributions of the means $\mon$ and $\moff$ 
can be selected only if the numbers of detected on-- and off--source 
events are positive. 
These prior distributions are classified by the rate parameters 
$\gp = \gq \to 1$, i.e. $\eta = \alal$ and $\rho = \alpha$, 
the shape parameters $\ssp =\ssq \to 0$, i.e. $p = \non > 0$ and 
$q = \noff > 0$. 
Hence, for the posterior distributions we have 
$\mon \sim \Gaa(\non,1)$ and 
$\mb \sim \Gaa(\noff,\alin)$, see~\app{App01}.
Then it follows from~\eqa{B13} that the non--negative difference 
of the on--source and background means, $\mon \ge \mb$, is 
\beql{B16}
\fdelp(x) = e^{-x} \ \frac{
\Ssum{i=0}{\non-1} {\non+\noff-i-2 \choose \non-i-1} 
\left( \alal \right)^{i} 
\frac{x^{i}}{i!}
}{
\Ssum{i=0}{\non-1} {\non+\noff-i-2 \choose \non-i-1} 
\left( \alal \right)^{i}
}, 
\qqc 
x \ge 0.
\eeq
The same result was presented in Ref.~\cite{Pro02}.

In our analysis, the Bayesian probability that a source is present
in the on--source region is given explicitly by, see~\eqa{B14},
\beql{B17}
\Pp = 
\left( \aoff \right)^{\noff}
\Ssum{i=0}{\non-1} {\non+\noff-i-2 \choose \non-i-1} 
\left( \aon \right)^{\non-i-1}.
\eeq

\subsubsection{Uniform priors}
\label{S02db}

Let us consider the uniform prior distributions of the means 
$\mon$ and $\moff$.
In such a case, the rate parameters $\gp = \gq \to 1$, i.e. 
$\eta = \alal$ and $\rho = \alpha$, the shape parameters 
$\ssp =\ssq = 1$, i.e. $p = \non+1$ and $q = \noff+1$.
The posterior distributions are 
$\mon \sim \Gaa(\non+1,1)$ 
and $\mb \sim \Gaa(\noff+1,\alin)$, see~\app{App01}.
Assuming the non--negative difference of the on--source 
and background means, $\mon \ge \mb$, we get from~\eqa{B13}
\beql{B18}
\fdelp(x) = e^{-x} \ \frac{
\Ssum{i=0}{\non} {\non+\noff-i \choose \non-i} 
\left( \alal \right)^{i} 
\frac{x^{i}}{i!}
}{
\Ssum{i=0}{\non} {\non+\noff-i \choose \non-i} 
\left( \alal \right)^{i}
}, 
\qqc 
x \ge 0.
\eeq
The same result was obtained in Refs.~\cite{Gre01,Pro01}.

The Bayesian probability that a source is present in the on--source 
region follows from~\eqa{B14}, namely,
\beql{B19}
\Pp = 
\left( \aoff \right)^{\noff+1}
\Ssum{i=0}{\non} {\non+\noff-i \choose \non-i} 
\left( \aon \right)^{\non-i}.
\eeq
We note that a quite different formula has been advocated 
in Ref.~\cite{Gre01}.
Its justification is based on the Bayes factor that accounts 
for a complex source model put against a simple background hypothesis.
However, as pointed out in Ref.~\cite{Gre01}, the significant
disadvantage is that the resultant probability strongly depends 
on the choice of the upper bound of the uniform prior used for 
the source activity.

Our Bayesian probabilities for the presence 
or absence of a source in the on--source region are easily obtained 
in the case of the null experiment, when no counts are registered 
in the on--source region, i.e. $\non = 0$ and $p = 1$, or in 
the experiment with zero background counts, i.e. $\noff = 0$ and $q = 1$,
\beql{B20}
\Pp_{\non = 0} = \left( \aoff \right)^{\noff+1}, \qqc
\Pm_{\noff = 0} = \left( \aon \right)^{\non+1}.
\eeq
Both these probabilities depend on the relationship between 
the on-- and off--source regions, on the on--off parameter $\alpha$.
Unlike other results~\cite{Gre01}, our approach provides us 
with well understandable solutions. 
For example, in the null experiment ($\non = 0$), the Bayesian probability 
of the source detection in the on--source region drops down with 
the increasing on--source exposure (increasing $\alpha$) as well as 
with the increasing number of registered off--source events.
If no events are registered at all,
we get $\Pp_{\non=\noff=0} = (1+\alpha)^{-1}$.

\subsubsection{Jeffreys'  priors}
\label{S02dc}

The key characteristic of the Jeffreys' prior distribution is that 
it is invariant under a transformation of parameters.
Thus, it expresses the same prior belief no matter which metric 
is used. 

In our notation scheme, Jeffreys' prior distributions of the on-- 
and off--source means are of the form introduced in~\app{App02}.
The posterior distributions are formally constructed if the rate 
and shape parameters of the Gamma distributions given in~\app{App01} 
satisfy $\ssp = \ssq = \frac{1}{2}$ and 
$\gp = \gq \to 1$, respectively.
Then, the distribution of the difference is obtained
putting $p = \non + \frac{1}{2}$, $q = \noff + \frac{1}{2}$, 
$\eta = \alal$ and $\rho = \alpha$ into the relevant equations. 

In particular, the distribution for the non--negative difference 
written in~\eqa{B12} implies the recent result based on Jeffreys' 
rule presented in Ref.~\cite{Kno01}.
Indeed, using the identities for the hypergeometric 
functions~\cite{Olv01}
\beql{B21}
\frac{a}{x^{a}} B_{x}(a,b) = {_{2}F_{1}}(a,1-b,a+1,x),
\eeq
and
\beql{B22}
{_{2}F_{1}}(a,b,c,x) = (1-x)^{a} {_{2}F_{1}}(a,c-b,c,\frac{x}{x-1}),
\eeq
one has, in our notation scheme,
\beql{B23}
q \alpha^{q} B_{\aoff}(q,p) = 
\left( \aon \right)^{q}{_{2}F_{1}}(q,1-p,q+1,\aoff) = 
{_{2}F_{1}}(q,p+q,q+1,-\alin).
\eeq
Substituting this result into~\eqa{B12} and decoding the values 
of the parameters $p$, $q$ and $\eta$, while $\gp = \gq \to 1$, 
the correspondence with the result written in Eq.(30) in 
Ref.~\cite{Kno01} is evident.

This consistency is due to the above mentioned invariant property 
of the Jeffreys' prior.
In our strategy, we started with the two independent variables 
$\mon > 0$ and $\moff > 0$ the prior distributions of which 
are given by Jeffreys' rule, see~\eqa{Z02.01} in~\app{App02}.
Choosing a new mean $\ms = \mon - \alpha \moff \ge 0$ and keeping
$\moff > 0$ unchanged, the bidimensional prior distribution 
considered in Eq.(15) in Ref.~\cite{Kno01} is easily obtained under 
this transformation.

It is worth stressing, however, that the probability of 
the absence of a source in the on--source zone that was derived 
in this study using the distribution of the difference
(see~\sct{S02c}) differs from the results of Refs.~\cite{Kno01,Cas01} 
when Jeffreys' rule for prior distributions is considered.
The reason is that the other methods do not benefit from all 
input information or do not fully utilize the Bayesian inference. 

In Ref.~\cite{Kno01}, the determination of the Bayesian probability 
of the background hypothesis was based on the questionable argument 
about how to choose the ratio of the arbitrary scale factors of 
the prior distributions of model parameters.
This ratio was derived following the {\it ad hoc} assumption 
that if no counts are observed in both zones, the probabilities 
of both the signal and background model remain the same.
However, one may successfully argue that such a null measurement 
with no background counts ($\non = \noff = 0$) 
should update our knowledge about the signal.  
The point is that one has additional information since the ratio 
of the on-- and off--source exposures is known by definition.
Therefore, the result of the null experiment with no background
counts is to prefer the signal alternative if $0 < \alpha < 1$ 
(larger off--source exposure) and vice versa.
Unfortunately, the premise behind the procedure that provides 
the probability of the background hypothesis, as advocated
in Ref.~\cite{Kno01}, does not take into account the possibility 
of different exposures.
Interestingly, while the probability of the no--source hypothesis $H_{0}$ 
is assumed to be $P(H_{0} \mid \non=\noff=0) = \frac{1}{2}$ in 
Ref.~\cite{Kno01}, we obtain from~\eqa{B08} for the Bayesian probability 
of the absence of a source in the on--source region a more intuitively 
appealing result
\beql{B24}
\Pm_{\non = \noff = 0} = I_{\aon} ( \frac{1}{2}, \frac{1}{2} ) = 
\frac{2}{\pi} {\rm arctan} ( \sqrt{ \alpha } ).
\eeq
With the increasing on--source exposure (increasing $\alpha$), 
the probability that a source is not in the on--source region 
increases if no counts ($\non = \noff = 0$) are detected 
in both on--off zones, for the uniform priors see~\eqa{B20}. 

In Ref.~\cite{Cas01}, 
a predictive distribution of background counts was utilized 
in order to assess to what extent the source model is not supported 
by the on--off data. 
Following Jeffreys' rule, the distribution for the background mean 
was modeled as the Gamma distribution with parameters deduced from 
the off--source observation using the method of moments. 
The significance of the signal deviation from the background
hypothesis was established based on the Poisson--Gamma 
mixture. 
In this approach, the on-- and off--source zones are treated 
differently.
The resultant $p$--values are to be interpreted as the probability 
of obtaining a result at least as extreme as the observed data if 
the null background hypothesis is true. 
Thus, such an approach does not fully exploit the Bayesian 
reasoning and, therefore, it cannot provide us with information 
what hypothesis is more likely, given the data.  

\subsubsection{Known background}
\label{S02dd}

The analysis may be adapted for the case of known background with 
remaining uncertainty in the on--source zone, for classical 
results see e.g. Ref.~\cite{Fel01}. 
Let us assume that the background mean $\mb$ is known, but we do not 
measure the counts due to the background during the experiment.
Such a situation may be reviewed as the limit $q \to \infty$ 
($q = \noff + \ssq$), $\alpha \to 0$ when $q \alpha = \mb$ remains 
a finite constant~\cite{Pro01}.
In our scheme, the difference of the two Poisson parameters enlarged 
by the constant background parameter follows the Gamma distribution, 
i.e. $\mon = (\delta + \mb) \sim \Gaa(p,\gp)$ where $p = \non + \ssp$ 
and $\gp > 1$ are parameters for the prior distribution of the on--source 
mean.
Therefore, the probability distribution of the difference $\delta$ 
is then given by
(here we have $\fdelb(x) = \fdelb( \delta=x \mid \non, \mb, \IIp)$)
\beql{B25}
\fdelb(x) = \frac{\gp^{p}}{\Gamma(p)} (x + \mb)^{p-1} e^{-\gp (x+\mb)},
\qqc
x \ge -\mb.
\eeq
In addition, assuming non--negative values of the difference $\delta$,
i.e. $\mon \ge \mb$, we have for its probability distribution
\beql{B26}
\fdelbp(x) = \frac{\fdelb(x)}{\Ppb} =
\frac{\gp^{p}}{\Gamma(p,\mb)} (x + \mb)^{p-1} e^{-\gp (x+\mb)}, \qqc
x \ge 0,
\eeq
where $\Gamma(a,x) = \Iint{x}{\infty} t^{a-1} e^{-t} \dif{t}$ 
is the upper incomplete Gamma function and
\beql{B27}
\Ppb = P(\delta > 0) = 
\Iint{0}{\infty} \fdelb(x) \dif{x} = 
\frac{\Gamma(p,\mb)}{\Gamma(p)},
\eeq 
is the probability of the presence of a source in the on--source 
region if the background mean is known.

Note that for the uniform prior distribution of 
the on--source parameter, 
when $p = \non+1$ ($\ssp = 1$) and $\gp \to 1$, 
the result written in~\eqa{B26} was obtained in Ref.~\cite{Hel02}.
More general expressions with the prior parameter $\gp \to 1$ 
were presented in Ref.~\cite{Pro01}.

\subsection{Source intensity}
\label{S02e}

With the complete information about the on--off measurement contained 
in the distribution of the difference $\delta$, we can estimate 
the source intensity.
We use the shortest credible interval $\langle \dell, \delr \rangle$ 
that includes the source intensity at a chosen significance level of $P$.
In order to obtain these intervals, one has to solve numerically
\beql{B28}
P = \Iint{\dell}{\delr} \fdel(x) \dif{x}, 
\qqc
\fdel(\dell) = \fdel(\delr),
\eeq
with the indicated condition on interval endpoints, 
if it can be fulfilled. 

In those cases when the lower endpoint of a credible interval is 
negative, an upper bound for the source intensity is usually required. 
Its value, $\delr$, is determined numerically using the integral 
in~\eqa{B28} where we put $\dell \to -\infty$ and relax 
the constraint on interval endpoints.

When it is known that a source may be present only in 
the on--source zone (see~\sct{S02d}), we obtain credible intervals 
for the source intensity by putting $\fdel(x) \to \fdelp(x)$ with 
$0 \le \dell$ into~\eqa{B28}. 
For upper bounds we set $\dell = 0$ while not using the constraint 
on interval endpoints. 

\section{Examples}
\label{S03}

\subsection{Source detection significance}
\label{S03a}

We present examples which illustrate some of the features 
of the method described in \sct{S02c} that allows us to assign 
the probability of the source absence or presence in the on--source 
zone for a pair of on--off measurements.
We focused on the cases with small numbers of events.
For this purpose, we use the significance derived from the Bayesian 
probability of the source presence in the on--source region using 
the standard normal variate, see~\sct{S02c}.
In each example we calculated Bayesian significances using 
the scale invariant, Jeffreys' as well as uniform prior distributions 
of the on-- and off--source means 
($\gamma = \gp = \gq \to 1$, $s = \ssp = \ssq$, $s \to 0$ or 
$s = \frac{1}{2}, 1$).
We also calculated the asymptotic Li--Ma significance~\cite{Lim01} 
with which, relying on the likelihood ratio method, the no--source 
hypothesis is rejected if it is true. 
We added a sign to the Li--Ma statistic $\SLM$ considering 
it as non--negative if $\non - \alpha \noff \ge 0$ and negative otherwise, 
i.e. $\SLM = \sgn(\non - \alpha \noff) \sqrt{\SLM^{2}}$, since 
the original statistic~\cite{Lim01} is equivalent to the absolute 
value of a standard normal variable.

\begin{figure}[ht!]
\wse
\includegraphics*[width=0.99\linewidth]{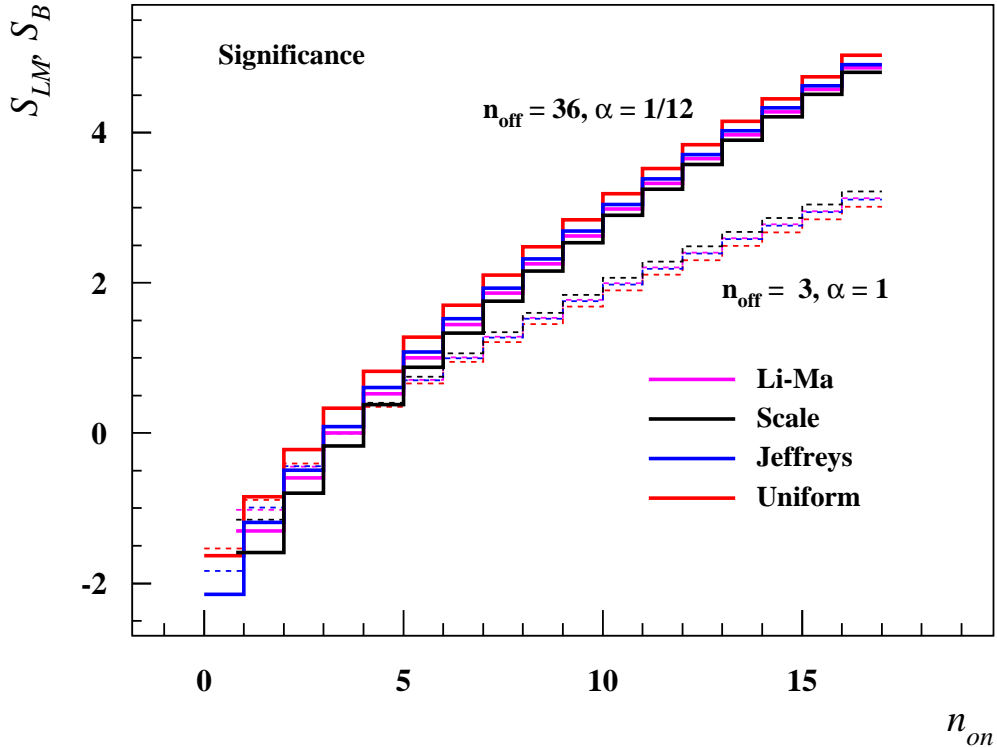}
\caption{\small
Significances for the source detection are shown as functions 
of the number of events detected in the on--source zone for 
two different off--source measurements.
In the first case (thick lines), $\noff = 36$ and 
$\alpha = \frac{1}{12}$.
In the second case (thin lines), $\noff = 3$ and $\alpha = 1$.
The Li--Ma significances are shown in magenta. 
The Bayes significances for scale invariant (black lines), Jeffreys' 
(blue lines) and uniform (red lines) prior distributions were derived 
from the probability of the source presence in the on--source region, 
see~\eqa{B09}.
}
\label{F01}
\end{figure}

In the first example, we chose the number of events detected 
in the off--source zone while varying the number of registered 
on--source counts.
We dealt with two cases.
In the first case, we assumed that $\noff = 36$ counts were 
detected in the off--source region the exposure of which is 
$12$--times larger than the exposure of the on--source zone, 
i.e. $\alpha = \frac{1}{12}$.
In the second case, we chose the same exposures of the on--
and off--source regions ($\alpha = 1$), and assumed that 
$\noff = 3$ events were registered in the off--source zone.
The numbers of on--source events were small, 
$\non \in \langle 0, 16 \rangle$.
Note that for $\non = 0$ the scale invariant and Li--Ma 
significances are not determined.
In~\fig{F01}, our results obtained within the Bayesian inference 
are compared with the asymptotic Li--Ma significances~\cite{Lim01}.
Obviously, better knowledge about background 
($\noff = 36, \alpha = \frac{1}{12}$) implies higher
absolute values of significances. 
Note that in this case ($\noff = 36, \alpha = \frac{1}{12}$), 
the Bayesian significances based on the uniform prior distributions 
are larger when compared with the scale invariant results since 
$\alpha \non < \noff = 36$, see~\sct{S02c}.
The opposite is true in the second case ($\noff = 3$, $\alpha = 1$) 
only when $\non > \noff = 3$.
The Bayesian significances based on Jeffreys' prior 
distributions always lie in between results derived assuming 
the uniform and scale invariant prior distributions, 
if the latter choice is possible ($\non > 0$ and $\noff > 0$).
\begin{figure}[ht!]
\wse
\includegraphics*[width=0.99\linewidth]{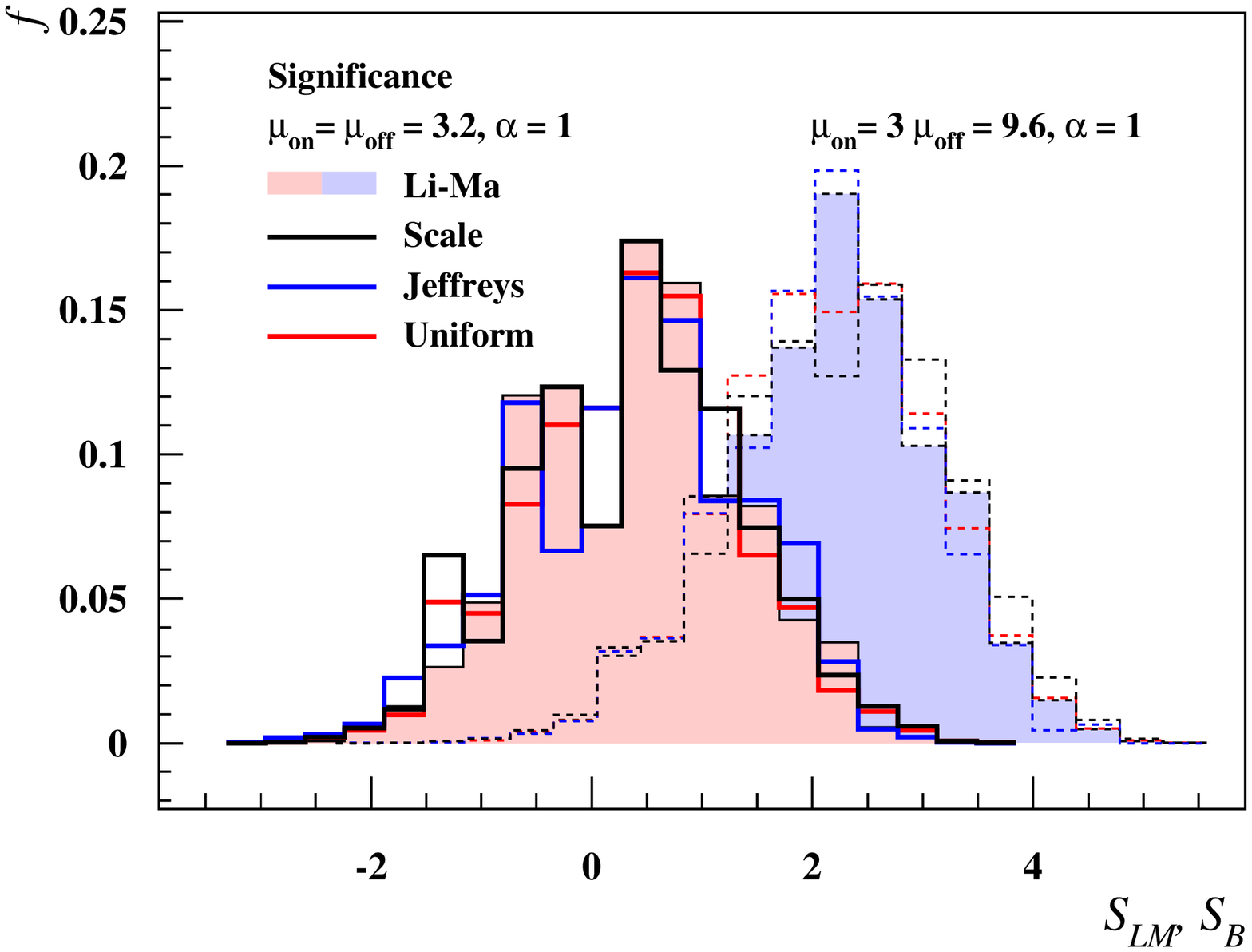}
\caption{\small
Distributions of significances for the source detection.
Histograms for the Li--Ma significances (filled areas) 
as well as for the Bayes significances using scale invariant 
(black lines), Jeffreys' (blue lines) and uniform (red lines) 
prior distributions are visualized.
Two examples for the same exposures of the on-- and 
off--source zones ($\alpha = 1$) are presented.
In the first example (light red area and thick lines), 
the on-- and off--source counts were generated with 
the mean parameters $\mon = \moff = 3.2$.
In the second example (light blue area and thin dashed lines), 
the mean parameters were $\mon = 3 \moff = 9.6$.
}
\label{F02}
\end{figure}

In the second example, using the Monte Carlo technique, 
we focused on the distributions of significances for the source
detection.
We generated $10^{5}$ pairs of on-- and off--source counts 
that follow the Poisson distribution with predefined source and 
background means, respectively, assuming that they were registered 
in the regions of the same exposures ($\alpha = 1$).
We determined the Bayesian probabilities of the source presence in 
the on--source region (see~\eqa{B09}) and the corresponding 
significances as well as the asymptotic Li--Ma significances~\cite{Lim01} 
for each pair of on-- and off--source counts.
In~\fig{F02}, we present the significance distributions for 
no source in the on--source zone, when the on-- and off--source 
means $\mon = \moff = 3.2$ (light red area and thick lines).
Typically, in about $10\%$ of on--off pairs, it happened that 
no events in the on-- or off--source zone were generated.
This results in dips in scale invariant (thick black line) and 
Li--Ma (light red area) histograms located around zero 
significance since, in such cases, the relevant significances 
cannot be determined.
In~\fig{F02}, we also show significance distributions that we received 
in the case when the source is present in the on--source region 
using $\mon = 3 \moff = 9.6$ (light blue and thin dashed lines).
In both presented cases, 
except the problem with zero counts,  
the significance distributions obtained using the Bayesian inference,
with the scale invariant, Jeffreys' or uniform prior distributions, 
are similar to each other as well as to the corresponding outputs 
obtained with the help of the asymptotic Li--Ma formula~\cite{Lim01}.

\subsection{Gamma--ray bursts}
\label{S03b}

The method described in~\sct{S02} was applied to the data
sets examined in Refs.~\cite{Kno01,Cas01}. 
We used information about very high energy (VHE) photons from 
gamma--ray bursts (GRB) collected by the VERITAS setup~\cite{Ver01} 
and by the Fermi Large Area Telescope~\cite{Fer01}.
These data sets of VHE photons detected during or shortly after 
12 bursts are listed in the first four columns in~\tab{T01}.
Typically, only a few VHE photons were registered in 
the directions of GRBs.
In most cases, the number of collected events is not too 
different from the corresponding number of events expected 
from background. 

We assumed the same prior distributions for the on-- and off--source
means with the common shape parameter, $s = \ssp = \ssq$, and zero 
rate parameters, i.e. $\gamma = \gp = \gq \to 1$.
With these restrictions we calculated the distributions 
of the difference of the on--source and background means.
The probability that a source is absent in the on--source region 
is then given by~\eqa{B08}.
We also determined credible intervals and, if appropriate, 
upper bounds of the source intensity at a given level of significance 
as described in~\sct{S02e}.

\begin{figure}[ht!]
\wse
\includegraphics*[width=0.99\linewidth]{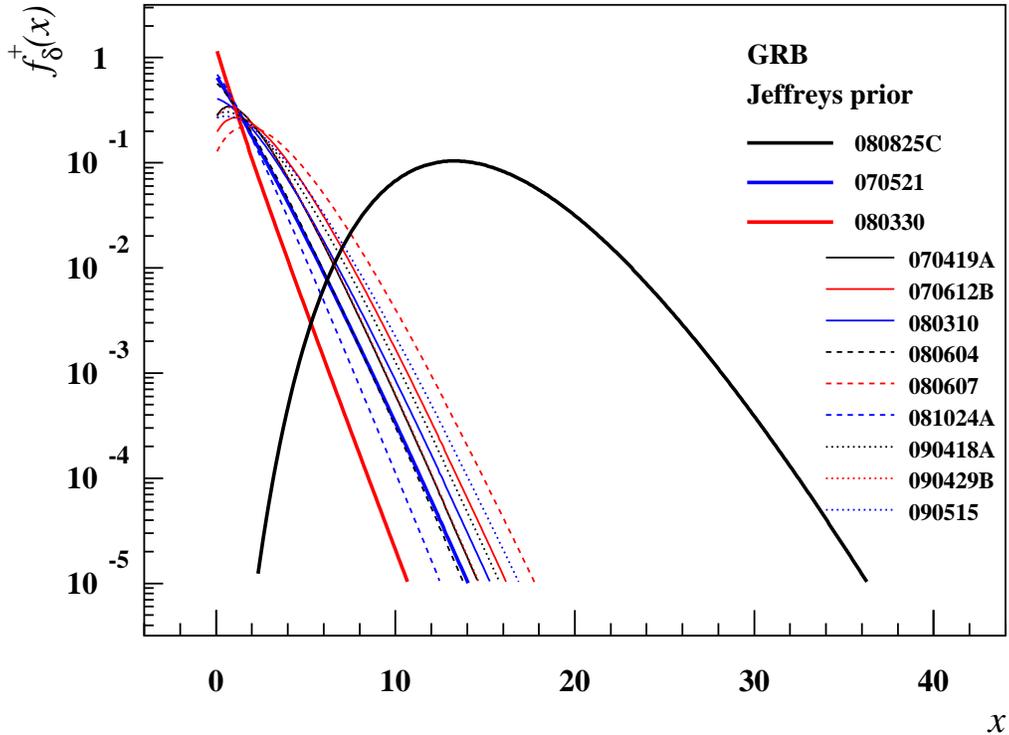}
\caption{\small
Distributions of the difference conditioned on non--negative values 
of the on--source rate for 12 data sets connected with GRB 
observations~\cite{Ver01,Fer01}. 
These results were obtained with Jeffreys' prior distributions for 
the on-- and off--source means ($s = \frac{1}{2}$ and $\gamma \to 1$). 
Full black, blue and red curves are for GRB~080825C, GRB~070521 and 
GRB~080330, respectively.
}
\label{F03}
\end{figure}

The conditional distributions of the difference 
($\mon \ge \mb$) for all data sets are depicted in~\fig{F03}.
These results were obtained with Jeffreys' prior distributions 
($s = \frac{1}{2}$ and $\gamma \to 1$). 
Specific properties of the GRB sources are summarized in~\tab{T01}.
In this table, we present the Bayesian probabilities of the absence 
of a source of VHE photons in the on--source zone ($\Pm$). 
Negative values of the corresponding Bayesian significance ($\SBa$) 
indicate that the absence of a source in the on--source region 
is more likely than its presence therein, i.e. $\Pm > 0.5$.
For all data sets, we also give credible intervals for the difference 
of the on--source and background means at a $99\%$ level 
of confidence.
With the distributions of the difference conditioned on 
the non--negative source intensity we reproduce the upper bounds 
($\delrr$) obtained in Ref.~\cite{Kno01} at the same level 
of confidence (see~\sct{S02dc}).
\begin{table}[ht!]
\vspace{-0.5cm}
\caption{\small 
GRB data collected by VERITAS~\cite{Ver01}, 
GRB~080825C was observed by Fermi--Lat~\cite{Fer01}.
The same sets of data as in Refs.~\cite{Kno01,Cas01} are used.
GRB assignments, measured counts and on--off parameters 
$\alpha$ are listed in the first four columns.
The Bayesian probability of the absence of a source in 
the on--source zone ($\Pm$), corresponding significance ($\SBa$) 
and Li--Ma significance ($\SLM$) are given in the following 
three columns.
We show credible intervals for the difference 
($\langle \dell , \delr \rangle$) 
and its upper bounds obtained by assuming that a source may 
be present only in the on--source zone ($\delrr$),
both at a $99\%$ level of confidence.
Confidence intervals for the difference 
($\langle \dell , \delr \rangle_{\rm L}$) 
derived in the unbounded profile likelihood method~\cite{Rol01} 
at the same level of confidence are given in the rightmost column.
For Bayesian results, Jeffreys' prior distributions 
($s=\frac{1}{2}$ and $\gamma \to 1$) were used. 
}
\begin{center}
{\footnotesize
\begin{tabular}{l r r r r r r r r r}
\\\hline\hline\\[-2mm]
GRB & $\non$  & $\noff$ & $\alpha$ & $P^{-}$ & $\SBa$ & $\SLM$ & 
$\langle \dell , \delr \rangle$ &  $\delrr$ & 
$\langle \dell , \delr \rangle_{\rm L}$ \\[2mm]
\hline\hline\\[-2mm]
070419A &  2 &  14 & 0.057 & 0.110 &  1.23 &  1.08 &  
-1.11, 7.74 & 6.88 & -0.88, 7.34 \\
070521  &  3 & 113 & 0.057 & 0.923 & -1.43 & -1.48 & 
-6.86, 2.91 & 6.12 & -6.77, 3.58 \\
070612B &  3 &  21 & 0.066 & 0.106 &  1.25 &  1.14 &  
-1.50, 8.61 & 8.00 & -1.23, 8.55 \\
080310  &  3 &  23 & 0.128 & 0.455 &  0.11 &  0.03 &  
-3.60, 6.92 & 7.16 & -3.37, 7.08 \\ 
080330  &  0 &  15 & 0.123 & 0.932 & -1.49 &       & 
-3.84, 3.43 & 4.10 & -3.38, 2.40 \\ 
080604  &  2 &  40 & 0.063 & 0.591 & -0.23 & -0.33 &  
-3.03, 5.10 & 6.12 & -2.93, 5.66 \\ 
080607  &  4 &  16 & 0.112 & 0.080 &  1.41 &  1.32 &  
-1.82,10.12 & 9.17 & -1.42, 9.84 \\ 
080825C & 15 &  19 & 0.063 & $7 \, 10^{-10}$ & 6.22 & 6.36 & 
5.05, 26.60 &  & 5.86, 26.12 \\
081024A &  1 &   7 & 0.142 & 0.441 &  0.15 &  0.01 &  
-2.13, 5.64 & 5.30 & -1.89, 5.19 \\ 
090418A &  3 &  16 & 0.123 & 0.233 &  0.73 &  0.64 & 
-2.50, 8.24 & 7.64 & -2.17, 8.01 \\ 
090429B &  2 &   7 & 0.106 & 0.106 &  1.25 &  1.12 & 
-1.04, 6.41 & 6.92 & -0.99, 7.41 \\ 
090515  &  4 &  24 & 0.126 & 0.282 &  0.58 &  0.50 & 
-3.25, 8.63 & 8.34 & -2.94, 8.66 \\[2mm] 
\hline\hline\\[-2mm]
\end{tabular}
}
\end{center}
\label{T01}
\end{table}
\begin{figure}[ht!]
\wse
\includegraphics*[width=0.99\linewidth]{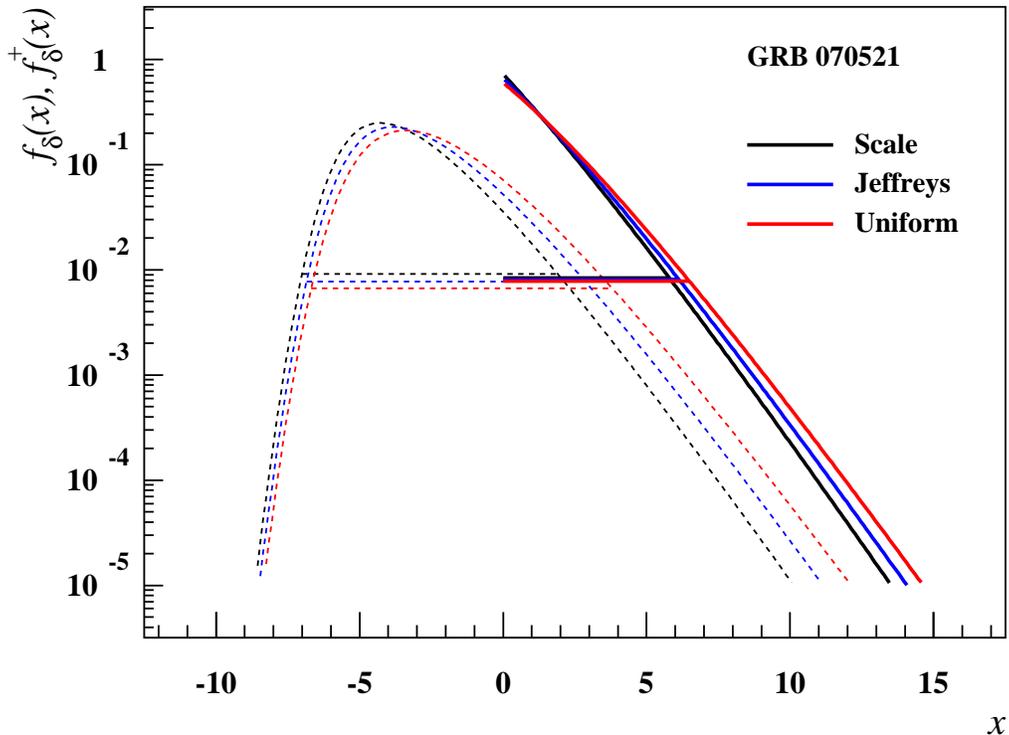}
\caption{\small
Distributions of the difference using the GRB~070521 data.
The distributions shown in black, blue and red were obtained 
with scale invariant ($s \to 0$), Jeffreys' ($s = \frac{1}{2}$) 
and uniform ($s = 1$) priors, respectively.
The thick full curves are for the distributions determined 
by assuming that a source may be present only in the on--source zone.
The dashed lines indicate solutions without this information.
Horizontal lines visualize credible intervals at a $99\%$ level 
of confidence.
}
\label{F04}
\end{figure}
\begin{figure}[ht!]
\wse
\includegraphics*[width=0.99\linewidth]{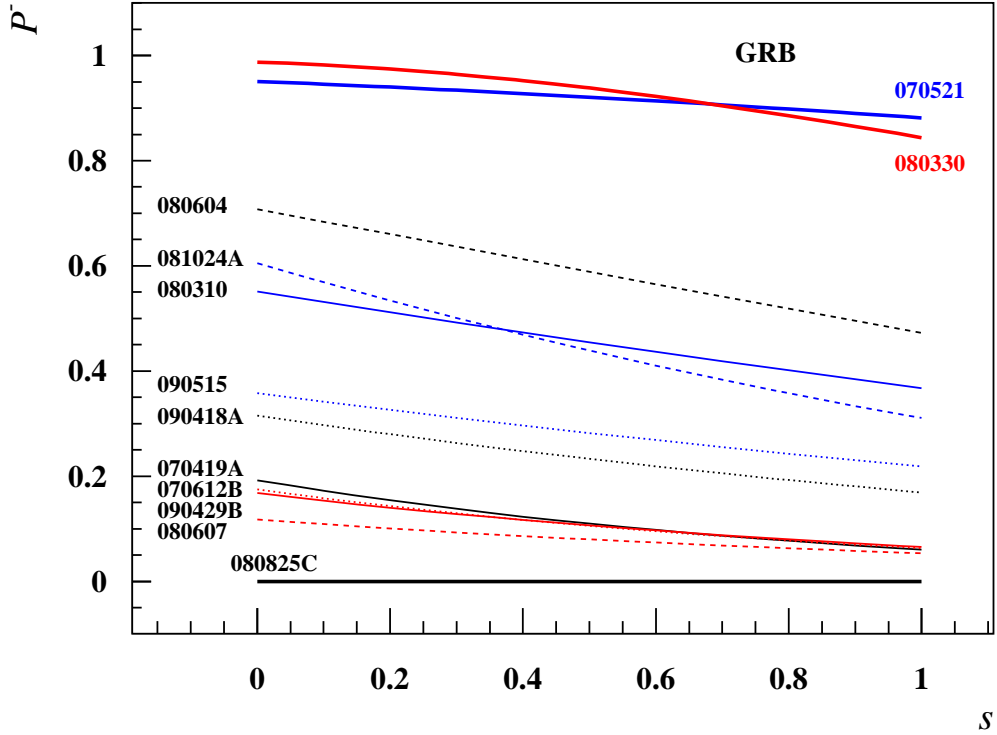}
\caption{\small
Probabilities of the source absence in the on--source zone
are shown as functions of the common shape parameter of prior
distributions. 
The thick full black, blue and red curves are for GRB~080825C, 
GRB~070521 and GRB~080330, respectively.
}
\label{F05}
\end{figure}
\begin{figure}[ht!]
\wse
\includegraphics*[width=0.99\linewidth]{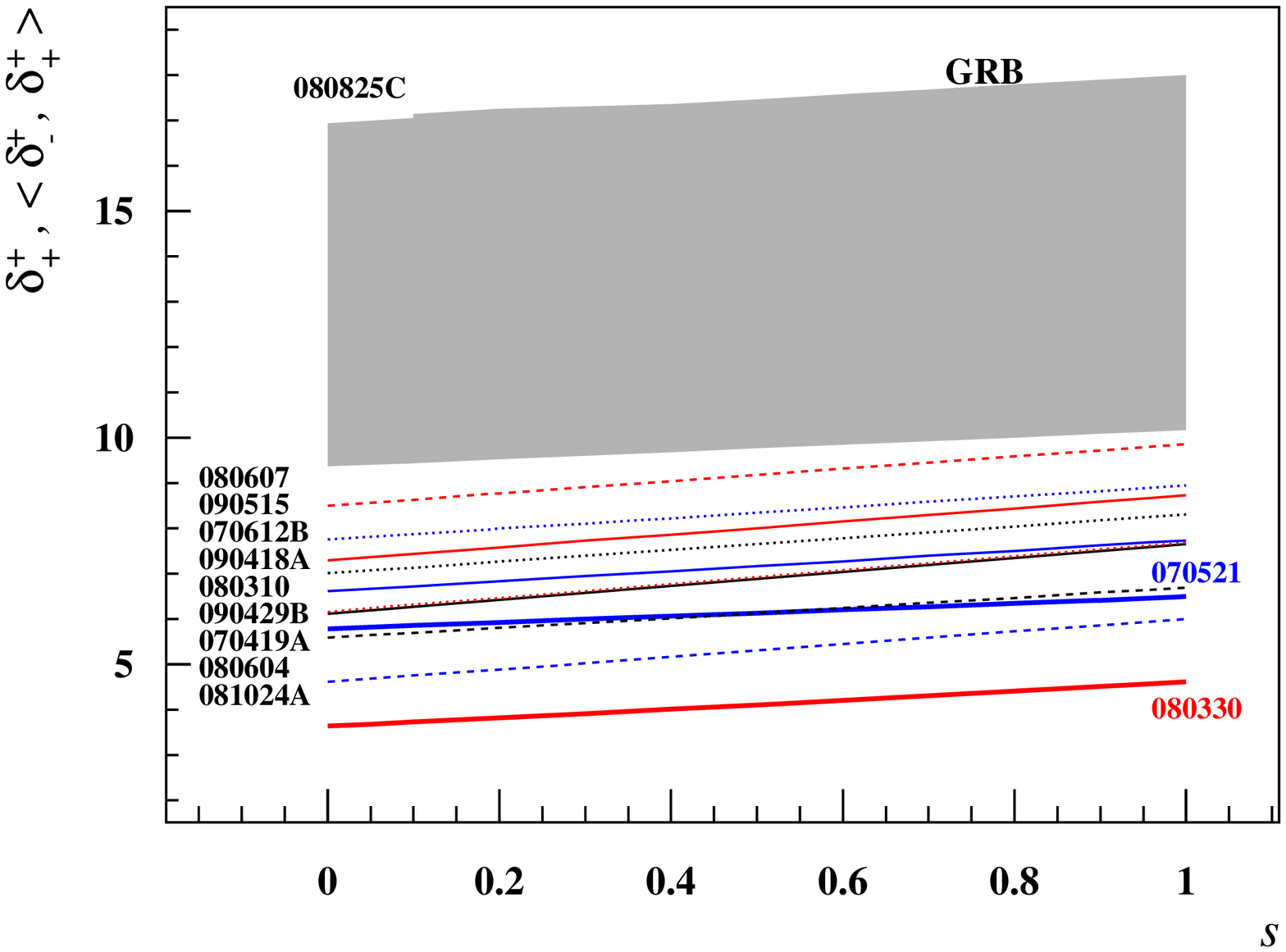}
\caption{\small
Upper bounds for the difference at a $99\%$ level of confidence 
are shown as functions of the common shape parameter of prior 
distributions. 
They were obtained by assuming that a source may be present only 
in the on--source zone.
The thick full blue and red curves are for GRB~070521 
and GRB~080330, respectively.
A grey band represents the $1\sigma$ credible interval for 
the activity associated with GRB~080825C.
}
\label{F06}
\end{figure}

In~\tab{T01}, we also present other on--off results which were obtained 
within the classical concept and have, therefore, a different meaning.
In particular, we calculated the Li--Ma significance~\cite{Lim01}.
Also confidence intervals derived using the unbounded likelihood 
method~\cite{Rol01} were determined.
It is worth stressing that both these classical statistics are obtained 
asymptotically relying upon the likelihood ratio and the Wilks' 
theorem~\cite{Wil01}.
The asymptotic confidence intervals or upper bounds are constructed 
in such way that they cover an unknown true value of the parameter 
under consideration with a specified probability.
The Li--Ma significance corresponds to the probability with which 
the null background hypothesis is rejected if it is true.  
We added a sign to the Li--Ma statistic $\SLM$
in order that $\SLM < 0$ if $\non - \alpha \noff <  0$.

There is a clear evidence that VHE photons from GRB~080825C were 
detected by the Fermi--Lat instrument~\cite{Fer01}.
The significance of the presence of a source in the on--source zone, 
$\SBa = 6.22$, provides the same conclusion as the asymptotic 
Li--Ma significance, as the original finding~\cite{Fer01} 
and other results~\cite{Kno01,Cas01}.
Our $1\sigma$ estimate of the source intensity, 
$\delta \in \langle 9.75, 17.49 \rangle$ obtained with 
Jeffreys' prior distributions, corresponds to previously 
presented estimates~\cite{Kno01,Cas01,Fer01}. 

Other data sets of VHE photons collected in the directions of GRBs
show no signature that would distinguish them from background data.
The Bayesian probabilities of the absence of a source in 
the selected on--source regions are above $8\%$, see~\tab{T01}.
The absolute values of the corresponding significance are 
below $1.50$.
Three data sets indicate a deficit of events in 
the on--source region, i.e. $\Pm > 0.5$ and $\SBa < 0$. 

Of particular interest are the results derived from the data 
associated with GRB~080330 since no on--source event
was recorded in this observation.
We recall that the data is easy to evaluate in the Bayesian 
approach.
No special assumptions or external constraints are needed.
The only exception is that the choice of the scale invariant 
priors ($s \to 0$) is excluded.
Using Jeffreys' prior distributions, our analysis yields 
the Bayesian probability of the absence of a source in 
the on--source zone of about $93\%$, see~\tab{T01}.
Our upper bound for the source intensity is somewhat higher
than the estimate which was obtained by extrapolation within 
the unbounded likelihood method~\cite{Rol01}.

With the aim to demonstrate the impact of different prior distributions, 
we choose the data set of GRB~070521. 
This data yields the lowest positive ratio of the number of 
on--source events with respect to the background counts expected 
in the on--source region.
In~\fig{F04}, various distribution functions of the difference 
for GRB~070521 are shown.
Three types of prior distributions were examined.
Namely, we present results based on the scale invariant ($s \to 0$, in black), 
Jeffreys' ($s=\frac{1}{2}$, blue) and uniform ($s=1$, red) prior 
distributions. 
The depicted distributions were obtained without (dashed curves) 
or with (full curves) 
assuming that a source may be present only in the on--source zone.
The former distributions are used in order to determine the probability 
of the source presence in the on--source zone.
The latter conditional distributions are then used for estimating 
credible intervals or upper bounds of the source intensity.

In this case, and also for other data sets listed in~\tab{T01}, 
we mostly obtained very similar results for the three types 
of prior distributions used for the on-- and off--source means.
This situation is documented in~\fig{F05} where the Bayesian 
probabilities for the source absence in the on--source zone 
are shown as functions of the common shape parameter of the prior 
distributions.
The no--source probabilities mostly slowly decrease with the increasing 
value of the prior shape parameter, from the scale invariant option 
($s \to 0$) down to the uniform choice ($s=1$).
The largest decline is found for the data associated with 
the observation of GRB~081024A when the lowest total number of events 
was detected.

Finally, the $99\%$ upper bounds of the source intensity, and 
the $1\sigma$ credible interval in case of 080825C, are depicted 
in~\fig{F06} as functions of the common shape parameter of 
the prior distributions. 
These characteristics were obtained by assuming that a source 
may be present only in the on--source region.
We learned that the limits of the source intensity are weakly 
dependent on the prior choice of the common shape parameter for
$0 < s \le 1$.   

\section{Conclusions}
\label{S04}

In this study we dealt the issue of detection of a source 
the activity of which is immersed in the surrounding background. 
For this purpose, we adopted the Bayesian concept that provides, 
on one side, a unified and intuitively appealing approach to 
the problem of drawing inferences from observations and, on the other 
side, it offers a powerful and sufficiently general framework 
for determining optimal behavior in the face of uncertainty.
As often reported, the Bayesian inference also allows us 
to alleviate some of the issues that affect conventional statistical 
approaches.

We have proposed a consistent description of the on--off measurement.
We focused on cases of small numbers of registered events that obey 
a Poisson distribution. 
For the on-- and off--source means, we used an adequately large 
class of conjugate prior distributions for the Poisson 
likelihood function.
It consists of Gamma distributions, each of which is parametrized 
by two parameters, by the rate and shape parameter.
The Gamma family includes several interesting and widely used options, 
i.e. scale invariant, uniform or Jeffreys' prior distributions.

We examined the distribution of the difference between
the on--source and background means.
This distribution is maximally noncommittal with regard to their dependence, 
but it carries all the information available from the on--off experiment.
Using it, the probability of the presence of a source in the on--source 
zone and other source properties are consistently derived within 
the Bayesian concept and, therefore, have well defined meaning.
We stress that our interpretation of the on--off data 
is different from reasoning behind hypothesis testing, regardless 
of whether the test is conducted in a classical or Bayesian framework.

The distribution of the difference is well suited for weak 
sources whose observations may reveal a signal either in the on--source 
or off--source zone, due to experimental limitations, for example.
Except one case, the proposed Bayesian solutions can be used 
for any number of on--off counts, including the null experiment 
or the experiment with no background.
To our knowledge, such results of the Bayesian inference have 
not yet been discussed in the literature.
By conditioning on the values of the difference we obtained 
a probability distribution that allows us to describe 
the on--off problem with a preassigned source in the on--source region 
the activity of which is to be examined. 
In this case, the resultant conditional distribution includes 
several results that have previously been obtained in other 
Bayesian approaches.
Using this distribution, well reasoned limits of the source activity 
are easily determined.

We also presented several numerical examples that may serve 
as guides for practical applications.
In most cases, when little is known about investigated phenomena, 
it turned out that the scale invariant, if applicable, 
or uniform prior distributions are good choices. 
The corresponding formulae reduce to simple algebraic sums, as described 
in~\sctw{S02da}{S02db} provided that a source may be present 
only in the on--source zone.
The Bayesian inference using Jeffreys' prior distributions should 
be a better compromise.
However, this option, as well as the choice of informative priors, 
requires more complicated calculations based on integral expressions.

\appendix

\section{Bayesian inference with Poisson likelihood}
\label{App01}
\setcounter{figure}{0}

We consider that the number of events registered in a counting experiment, 
a random variable $n$, obeys the Poisson distribution with a mean 
$\mu > 0$, i.e. $n \sim \Poo(\mu)$. 
The probability to observe $n$ events ($n = 0,1, \dots$) is 
\beql{Z01.01}
\PPo(n \mid \mu) = \frac{\mu^{n}}{n!} e^{-\mu}.
\eeq
Our aim is to deduce some information about the Poisson mean
$\mu$ from a measurement in which $n$ counts were registered.
For this purpose, we adopt the Bayesian reasoning.
The probability distribution of the Poisson mean to have 
the value $\mu > 0$ is found by means of Bayes' theorem
\beql{Z01.02}
f(\mu \mid n) \propto L(\mu \mid n) p(\mu),
\eeq
where $L(\mu \mid n) = \PPo(n \mid \mu)$ is the likelihood function
and $p(\mu)$ denotes the prior distribution of the mean $\mu$.

The problem is solved once we specify the form of the prior 
distribution.
To this end, we use Gamma distributions that provide 
a family of conjugate prior distributions for the Poisson 
likelihood
function
\beql{Z01.03}
p(\mu) = \fGa(\mu \mid s, \lambda) = 
\frac{\lambda^{s}}{\Gamma(s)} \mu^{s-1} e^{-\lambda \mu}, 
\eeq
where $s > 0$ is the shape parameter, $\lambda > 0$ denotes 
the rate parameter and $\Gamma(s)$ is the Gamma function. 
Notice that the mean and variance of a random variable obeying 
the Gamma distribution are 
${\rm E} (\mu) = \lambda^{-1} s$ and 
${\rm Var} (\mu) = \lambda^{-2} s$, 
respectively.
Hence, with the increasing value of the shape parameter $s$,
the prior distribution is peaked at larger values around a mode 
$\lambda^{-1} (s-1)$.
With the increasing value of the rate parameter $\lambda$, 
that shifts the position of the mean towards lower values, 
the prior distribution becomes narrower.

The Gamma family of prior distributions is sufficiently large.
The two prior parameters $s$ and $\lambda$ may be chosen 
to contain our degree of belief about the problem before 
the experiment is conducted.
Notice that traditionally accepted prior assumptions about the studied 
parameter are included among these possibilities. 
For example, in a limiting case, if $\lambda \to 0$, the choice
$s = 1$ represents the uniform prior, 
$s = \frac{1}{2}$ is for the Jeffreys' prior (see~\app{App02}) and, 
if $n > 0$, then the scale invariant prior distribution with $s \to 0$ 
may be selected.

The posterior distribution of the Poisson mean $\mu$ then depends 
on the prior choice and experimental data.
If $n$ events were collected, one easily finds that 
$\mu \sim \Gaa(p,\gamma)$, where $p=n+s > 0$ and $\gamma = \lambda + 1 > 1$, 
follow from~\eqa{Z01.02} for the prior distributions chosen from
the Gamma family defined in~\eqa{Z01.03}.
Hence, the posterior distribution function is
\beql{Z01.04}
f(\mu \mid n) = \fGa(\mu \mid p, \gamma) = 
\frac{\gamma^{p}}{\Gamma(p)} \mu^{p-1} e^{-\gamma \mu}.
\eeq
Let us finally note, that for the random variable $\mu' = k \mu$, 
where $k > 0$ is a constant, one obtains 
$\mu' \sim \Gaa(p,\frac{\gamma}{k})$.

\section{Jeffreys' prior}
\label{App02}
\setcounter{figure}{0}

By definition, the Jeffreys' prior is proportional to the square root of 
the determinant of the Fisher information. 
In the case of a single--valued Poisson mean $\mu > 0$,
it is written
\beql{Z02.01}
p(\mu) \propto \sqrt{ -{\rm E} \left[ 
\frac{\partial^{2} \ln L(\mu \mid n)}{\partial^{2} \mu} \right] } =
\frac{1}{\sqrt{\mu}},
\eeq
where $L(\mu \mid n) = \PPo(n \mid \mu)$ is the likelihood function
given in~\eqa{Z01.01} and ${\rm E}$ denotes the mean value with respect 
to the Poisson model under study.

\vspace*{0.5cm}
{\bf Acknowledgment:}
We would like to thank the two unknown reviewers for their 
constructive criticism and many useful suggestions that helped us 
with the presentation of our study.
This work was supported by the Czech Science Foundation under 
project~14-17501S. 

\vspace*{0.5cm}






\end{document}